\documentclass[10pt,conference]{IEEEtran}

\usepackage{amsmath}
\usepackage{amssymb}
\usepackage{graphics}
\newtheorem{Theorem}{Theorem}

\begin{document}

\title{
Encoding via Gr\"obner bases and discrete Fourier transforms for several types of algebraic codes}

\author{
\authorblockN{Hajime Matsui}
\authorblockA{Dept.\ of Electronics and Information Science\\
Toyota Technological Institute\\
Hisakata 2-12-1, Tenpaku, Nagoya 468-8511, Japan\\
hmatsui@toyota-ti.ac.jp}
\and
\authorblockN{Seiichi Mita}
\authorblockA{Dept.\ of Electronics and Information Science\\
Toyota Technological Institute\\
Hisakata 2-12-1, Tenpaku, Nagoya 468-8511, Japan\\
smita@toyota-ti.ac.jp}
}

\maketitle

\begin{abstract}
We propose a novel encoding scheme for algebraic codes such as 
codes on algebraic curves, multidimensional cyclic codes, and hyperbolic 
cascaded Reed--Solomon codes and present numerical examples. 
We employ the recurrence from the Gr\"obner basis of the locator ideal for a set of rational points
and the two-dimensional inverse discrete Fourier transform.
We generalize 
the functioning of the generator polynomial for Reed--Solomon codes and 
develop systematic encoding for various algebraic codes.
\end{abstract}

\section{Introduction}
Heretofore, there have been some researches on the encoding of codes on algebraic 
curves, although they are fewer than researches on the decoding of codes.
Heegard \textit{et al.} \cite{Heegard} proposed an encoding for 
linear codes with nontrivial automorphism groups by using Gr\"obner bases for modules over polynomial rings, 
which was applied by Chen {\it et al.} \cite{Chen-Lu}.
Matsumoto {\it et al.} \cite{Matsumoto-Oishi-Sakaniwa} proposed another 
encoding for codes on curves, 
based on the linear combination of extended Reed--Solomon (RS) codes
by the work of Yaghoobian {\it et al.} \cite{Yaghoobian}.

In this research, we propose a novel encoding scheme for various algebraic 
codes; this scheme is considered to be the natural generalization of the well-known encoding for RS codes.
We first establish a simple but non-systematic encoding that employs 
two-dimensional (2-D) inverse discrete Fourier transforms (IDFT) and that
generalizes the encoding for RS codes by using one-dimensional IDFT (that is, the Mattson--Solomon polynomial).
Since the syndromes correspond to the discrete Fourier transform (DFT), we 
also obtain a concise decoding via Berlekamp--Massey--Sakata (BMS) 
algorithm.
Next, we establish systematic encoding in the sense of the separation of 
given information and generated redundant in a resulting code-word.
This second method of encoding employs a Gr\"obner basis and its 2-D linear 
feedback shift-register and corresponds to the Euclidean division by the 
generator polynomial in the case of RS codes.

Both the methods often employ the enlargement of the finite-field arrays to 
the entire plane by the elements of Gr\"obner bases, typically, the defining 
equation of the algebraic curves.
As a more essential idea of our encoding and decoding scheme,
we can mention the following duality for substitution
$$(x^{i}y^{j})(\alpha^{r},\alpha^{s})
=(x^{r}y^{s})(\alpha^{i},\alpha^{j})=\alpha^{ir+js}.$$
Then, the rational points having any zero are exceptional;
however, they can be treated similarly to the case of lengthened RS codes as shown in section \ref{Treatment}.

\section{Codes on algebraic curves}\label{Codes on algebraic curves}

Let $\mathbb{Z}_{0}$ denote the set of non-negative integers.
Let $\mathcal{X}$ denote a non-singular C${}_{a}^{b}$ algebraic curves over 
$K:=\mathbb{F}_{q}$ for $a,b\in\mathbb{Z}_{0}$ with $a<b$ and gcd$(a,b)=1$.
Then, the genus of $\mathcal{X}$ is given by $g:=(a-1)(b-1)/2$,
and $\mathcal{X}$ has only one $K$-rational point at infinity $P_{\infty}$.
We fix a primitive element $\alpha$ of $K$.
Let $\mathcal{P}=\{P_{h}\}_{0\le h<n}$ denote a set of $K$-rational points of 
the form $P_{h}=(\alpha^{r},\alpha^{s})$, i.e., non-zero coordinates.
We construct codes of symbol-field $K$ on $P_{h}$'s in $\mathcal{P}$;
$K$-rational points whose coordinates include zero are considered in 
section \ref{Treatment}.
We define a subset $\Phi_{m}$ of $\mathbb{Z}_{0}^{2}$ as
$$\Phi_{m}:=\{(i,j)\in\mathbb{Z}_{0}^{2}\,|\,i<q-1,\,j<a,\,
ai+bj\le m\},$$
where $ai+bj$ is equal to the pole order $o(x^{i}y^{j})$ of $x^{i}y^{j}$ at 
$P_{\infty}$.
In this study, we consider codes on algebraic curves
\begin{equation}\label{C(m)}
\mathcal{C}(m):=\left\{
(c_{h})\in K^{n}\left|\,
c(\alpha^{i},\alpha^{j})=0,\;(i,j)\in\Phi_{m}
\right.\right\},
\end{equation}
where $c(x,y):=\sum_{h=0}^{n-1}c_{h}z_{h}$ with monomials 
$z_{h}:=x^{r}y^{s}$ for $P_{h}=(\alpha^{r},\alpha^{s})$.
For simplicity, we assume $m>2g-2$;
then, we obtain $n-k=m-g+1=\sharp\Phi_{m}$.

{\it Elementary encoding:}
The condition $\{c(\alpha^{i},\alpha^{j})=0\}$ in \eqref{C(m)} is equivalent 
to the ordinary linear system
\begin{figure*}[!t]
\centering
  \resizebox{15cm}{!}{\includegraphics{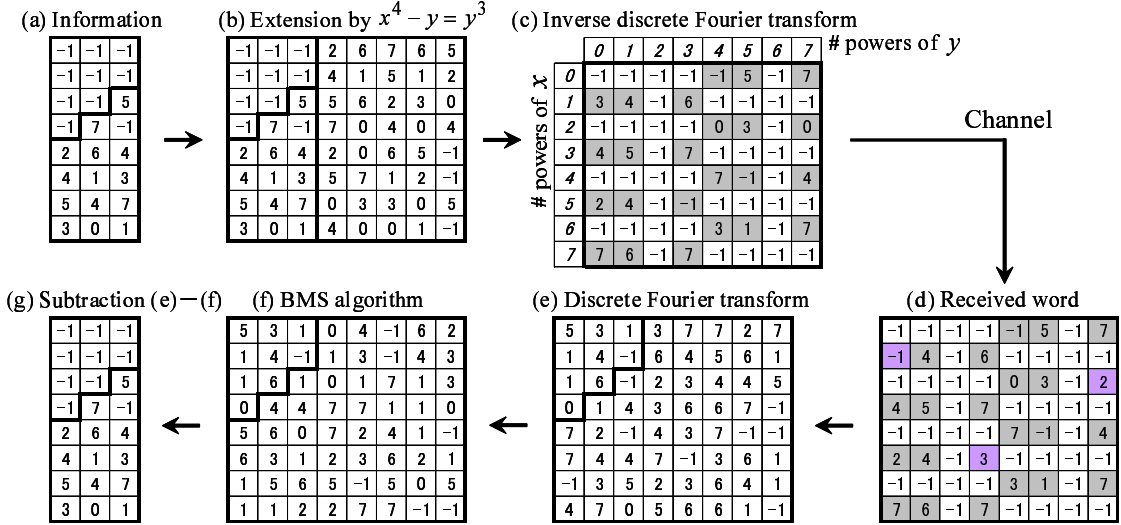}}
\caption{Flow chart of a non-systematic encoding by IDFT and a decoding by BMS algorithm with DFT for Hermitian code $\mathcal{C}(11)$ over GF($3^2$); the shaded values in (c) and (d) indicate the values on the $K$-rational points with non-zero coordinates. Array (c) represents a code-word and array (g) indicates that three errors have been corrected.\label{Fourier encoding}}
\vspace{-2mm}
\end{figure*}
\begin{equation}\label{linear system}
(c_{h})_{0\le h<n}\left[
z_{h}(Q_{l})\right]_{0\le h<n,\,0\le l<n-k}
=\mathbf{0},
\end{equation}
where $Q_{l}:=(\alpha^{i},\alpha^{j})$ for $(i,j)\in\Phi_{m}$ with order 
$l\le l'\Leftrightarrow ai+bj\le ai'+bj'$.
An encoding method for $\mathcal{C}(m)$ is the use of the generator matrix 
$G:=\left[\begin{array}{c|c}E_{k}&-H\end{array}\right]$, where $E_{k}$ is 
the $(k\times k)$ identity matrix and $H$ is obtained from
$\left[\begin{array}{c}H\\\hline E_{n-k}\end{array}\right]$ by the row 
transform of $\left[z_{h}(Q_{l})\right]$ and, if needed, the order-changing 
of $\mathcal{P}$.
Then, we can systematically encode information symbols $(I_{\kappa})_{0\le\kappa<k}$ to a 
code-word $(c_{h}):=(I_{\kappa})G$.
However, this requires the multiplication of ($k\times(n-k)$) matrix $H$.
Thus, our goal should be to eliminate the matrix-multiplication from the encoding algorithm.

On the other hand, with regard to the computing of syndromes, the situation is similar 
but better than the above encoding because of \eqref{C(m)}.
We suppose that an error-vector $(e_{h})$ has occurred during the transmission 
of $(c_{h})$ and we have received a word $(r_{h}):=(c_{h})+(e_{h})$.
Then, the syndrome decoding requires the $(n-k)$ values of syndrome
$(r_{h})\left[z_{h}(Q_{l})\right]$, which agree with
$\left(r(Q_{l})\right)$ for our expression of $\mathcal{C}(m)$.
This generalizes the syndrome-calculation for RS codes by the substitution of 
the roots of the generator polynomial.
Hence, we consider an effective encoding method based on our definition 
\eqref{C(m)} of $\mathcal{C}(m)$.

\section{Encoding by 2-D discrete Fourier transform}\label{Encoding 2-D}
In this section, we provide the example of a Hermitian code over 
$K:=\mathbb{F}_{9}$ with defining equation $y^{3}+y=x^{4}$ of genus $g=3$, 
the minimal pole order (first non-gap) $a=3$, and 24 $K$-rational points of 
$xy\not=0$ and finite. The primitive element $\alpha$ is fixed to satisfy 
$\alpha^{3}+\alpha+1=0$, and the non-zero element $\alpha^{i}$ ($0\le i<8$) is 
simply denoted as $i$ (resp. zero as $-1$). Note that 
$-\alpha^{0}=\alpha^{4}\not=\alpha^{0}$.
We represent 24 $\mathbb{F}_{9}$-rational points as monomials 
$\{x^{r}y^{s}\,|\,(\alpha^{r},\alpha^{s})\in\mathcal{P}\}$, which correspond 
to the shaded boxes of (c) in Fig. \ref{Fourier encoding}.

Let $\Phi\subset\mathbb{Z}_{0}^{2}$ be the support of a Gr\"obner basis of 
the ideal $I_\mathcal{P}:=\{f\in 
K[\mathcal{X}]\,|\,f(P_{h})=0,\,P_{h}\in\mathcal{P}\}$, i.e., $\Phi$ 
corresponds to the set of monomial representatives of 
$K[\mathcal{X}]/I_\mathcal{P}$, where 
$K[\mathcal{X}]=K[x,y]/(y^{3}+y-x^{4})$. Then, we have 
$\sharp\Phi=\sharp\mathcal{P}$. For Hermitian codes on non-zero coordinates, 
$\Phi$ agrees with $\{(i,j)\in\mathbb{Z}_{0}^{2}\,|\,i<q-1,\,j<a\}$;
in general, $\Phi$ is its subset.
We arrange the information symbols $(I_{(i,j)})$ on $\Phi\backslash\Phi_{m}$, 
and then obtain $(I_{(i,j)})_{(i,j)\in\Phi}$ by considering $I_{(i,j)}:=0$ if 
$(i,j)\in\Phi_{m}$, as (a) in Fig. \ref{Fourier encoding}.
Furthermore, the Gr\"obner basis of $I_\mathcal{P}$ extends 
$(I_{(i,j)})_{(i,j)\in\Phi}$ into $(I_{(i,j)})$ for 
$(i,j)\in\mathbb{Z}_{0}^{2}$ with $0\le i,j<q-1$; for Hermitian codes, 
$I_{(i+a+1,j-a)}-I_{(i,j-a+1)}=:I_{(i,j)}$ from the defining equation, as 
shown in Fig. \ref{Fourier encoding}(b), where $i:=i\,\mathrm{mod}\,(q-1)$ if $i\ge q-1$.

To encode $(I_{(i,j)})$, we perform the 2-D IDFT for $(I_{(i,j)})$:
$$
c_{(r,s)}:=\hspace{-3mm}\sum_{0\le i,j<q-1}\hspace{-3mm}
I_{(i,j)}\alpha^{-ri-sj}
=\hspace{-3mm}\sum_{0\le i,j<q-1}\hspace{-3mm}
I_{(q-1-i,q-1-j)}\alpha^{ri+sj}.
$$
Then, by substituting $P_{h}=(\alpha^{r},\alpha^{s})$ into $I(x,y)$, we have
\begin{align}
c_{h}&:=c_{(r,s)}=\sum_{0\le i,j<q-1}\label{DFT}
I_{(i,j)}x(P_{h})^{-i}y(P_{h})^{-j}\\
&=I(P_{h})\;\;\mathrm{for}\;\;
I(x,y):=\hspace{-2mm}\sum_{0\le i,j<q-1}\hspace{-2mm}
I_{(q-1-i,q-1-j)}x^{i}y^{j}.\nonumber
\end{align}

\begin{Theorem}\label{lemma of Fourier}
We have $c_{(r,s)}=0$ if there is no $P_{h}\in\mathcal{P}$ with 
$P_{h}=(\alpha^{r},\alpha^{s})$.
Moreover, the transform \eqref{DFT} defines an injective linear map and a 
code-word $(c_{h})_{0\le h<n}\in\mathcal{C}(m)$.
\end{Theorem}

\begin{figure*}[!t]
\centering
  \resizebox{15cm}{!}{\includegraphics{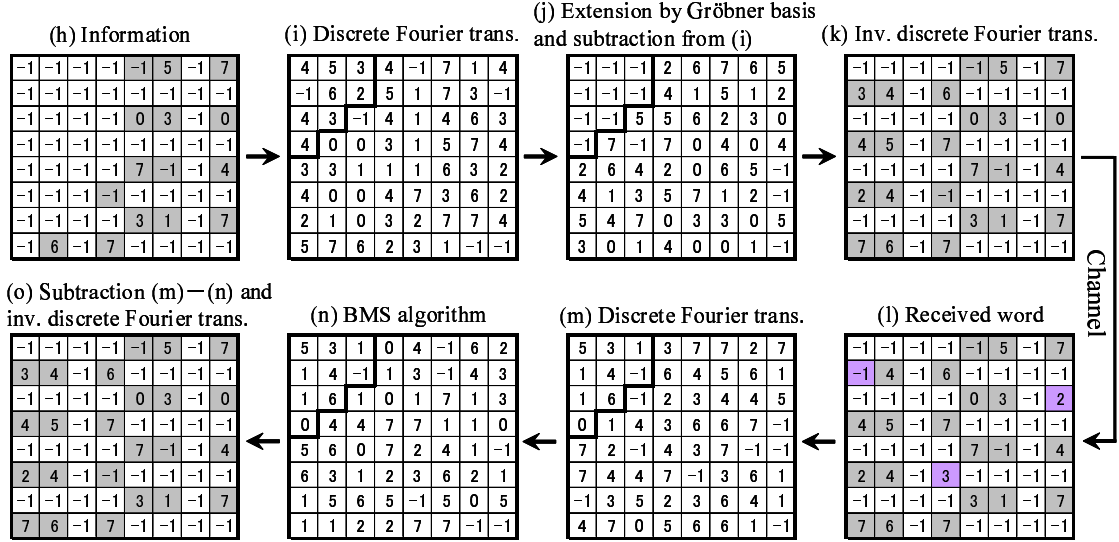}}
\caption{Flow chart of systematic encoding by Gr\"obner basis, decoding by BMS algorithm with DFT for code $\mathcal{C}(11)$; array (k) represents a systematic code-word and array (o) indicates that the correct information has been obtained.\label{systematic}}
\vspace{-3mm}
\end{figure*}

We omit the proof and discuss RS-code case in section \ref{Motivation}.

The received word $(r_{h})_{0\le h<n}$ is viewed as $(r_{(i,j)})$ for $0\le 
i,j<q-1$ and $r_{(i,j)}:=r_{h}$ if there is $P_{h}\in\mathcal{P}$ with 
$P_{h}=(\alpha^{i},\alpha^{j})$; otherwise $r_{(i,j)}:=0$ (cf. Fig. \ref{Fourier encoding}(d)).

The syndromes from the received word $(r_{h})$ can be obtained by the 
substitution of $(\alpha^{i},\alpha^{j})$ for $(i,j)\in\Phi_{m}$ into 
$r(x,y)$,
as described in Section \ref{Codes on algebraic curves}.
In our framework, it is convenient, as shown in Fig. \ref{Fourier encoding}(e), to substitute the entire $\{(i,j)\}_{0\le i,j<q-1}$, which can be considered as the DFT $\left(r(\alpha^{i},\alpha^{j})\right)_{0\le i,j<q-1}$.
Then, we have $r(\alpha^{i},\alpha^{j})=e(\alpha^{i},\alpha^{j})+I_{(i,j)}$ 
through the error polynomial $e(x,y):=\sum_{h=0}^{n-1}e_{h}z_{h}$ and the 
extended information symbols $(I_{(i,j)})$ because of our encoding and the 
2-D Fourier inversion formula
$$\sum_{0\le r,s<q-1}\sum_{0\le i,j<q-1}I_{(i,j)}\alpha^{-ri-sj+ri'+sj'}
=(q-1)^{2}I_{(i',j')}.$$

We notice that the values $(e(\alpha^{i},\alpha^{j}))$ are not yet known 
for $(i,j)\in\Phi\backslash\Phi_{m}$ since $I_{(i,j)}\not=0$ outside 
$\Phi_{m}$.
To obtain and subtract all syndrome-values $(e(\alpha^{i},\alpha^{j}))$ for 
$0\le i,j<q-1$ from $(r(\alpha^{i},\alpha^{j}))$, we run the BMS 
algorithm to calculate the Gr\"obner basis of the ideal $I_\mathcal{E}$, 
where $\mathcal{E}$ denotes the set of error locations.
Since the Gr\"obner basis 
provides the 2-D linear recurrence formula for syndromes, we can extend 
$(e(\alpha^{i},\alpha^{j}))_{(i,j)\in\Phi_{m}}$ to the entire plane,
where the array (f) represents the result.
Finally, as illustrated in Fig. \ref{Fourier encoding}(g),
the information $(I_{(i,j)})_{(i,j)\in\Phi\backslash\Phi_{m}}$
(and its extension $(I_{(i,j)})_{0\le 
i,j<q-1}$) is obtained by (e) minus (f).

Thus, DFT is utilized for both the encoding and computing of syndromes;
the decoding consists of two steps, i.e., this DFT and BMS 
algorithm to remove the syndrome-values, without Chien search and 
error-evaluator formula.

\section{Systematic encoding}\label{Systematic encoding}

Since the conventional RS codes are usually encoded systematically, it is 
natural to consider effective systematic encoding for codes on algebraic 
curves. 
However, while the roots of the generator polynomial can be considered 
a subset of locations in RS code-words, it does not hold in general for our 
$\mathcal{C}(m)$ since actually $\{Q_{l}\}_{0\le 
l<n-k}\not\subset\mathcal{P}$.
In this section, we apply Theorem \ref{lemma of Fourier} and its argument
to this problem and obtain a satisfactory solution.

As preliminaries, we choose $\wp$ and $\wp'$ so that 
$\wp\cup\wp'=\mathcal{P}$, $\wp\cap\wp'=\emptyset$, and $\sharp\wp=n-k$; 
$\wp$ is the redundant-point set and is considered to satisfy
$\wp=\{P_{h}\}_{0\le h<n-k}$
without loss of generality, and $\wp'$ is the information-point set. 
We calculate the Gr\"obner basis of the ideal $I_{\wp}$ in advance.
In the example of Fig. \ref{systematic},
the shaded boxes in (h) indicate $\wp'$, 
and we obtain the Gr\"obner basis by the BMS algorithm as follows:
$$
\hspace{7.5mm}
\resizebox{6.8cm}{!}{\includegraphics{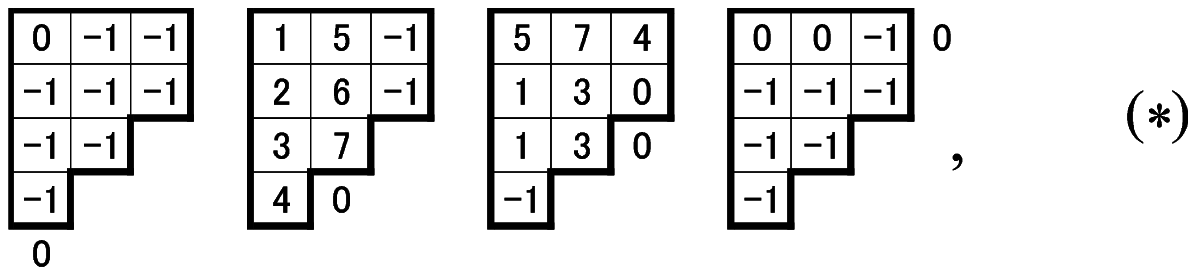}}
\vspace{-3mm}
$$
where, for example, the leftmost array represents the polynomial $1+x^{4}$.
For simplicity, we assume that
the support of the Gr\"obner basis for $I_{\wp}$
is generic \cite{ISITA04}, that is, the support corresponds to 
$L(mP_{\infty})$; this assumption is not very strong
since the generic support has the probability $(q-1)/q$.

Then we represent $k$ information symbols $\{I_{(i,j)}\}_{(i,j)\in\wp'}$
as shown in Fig. \ref{systematic}(h). 
To generate the redundant part of the code-word, we first compute its DFT 
$\{\tilde{I}_{(i,j)}\}_{0\le i,j<q-1}$ by
\begin{equation}\label{inf}
\tilde{I}_{(i,j)}:=I(\alpha^{i},\alpha^{j})\;\;\mathrm{for}\;\;I(x,y):=\sum_{(\alpha^{i},\alpha^{j})\in\wp'}I_{(i,j)}x^{i}y^{j},
\end{equation}
and then, we extend $\{\tilde{I}_{(i,j)}\}_{(i,j)\in\Phi_{m}}$ (nine values 
in Fig. \ref{systematic}(i)) on the support into $\{\breve{I}_{(i,j)}\}_{0\le i,j<q-1}$ on the entire plane by the Gr\"obner basis ($\ast$),
or more precisely, by its recursive formula \eqref{generation} in section
\ref{Proof of Theorem}.
If we perform IDFT for the negative $\{-\breve{I}_{(i,j)}\}$ of the extended 
array, the redundant part can be obtained since the resulting values on 
$\wp'$ are zero by Theorem \ref{lemma of Fourier} and their DFT (i.e., syndrome) agrees with $\{-\tilde{I}_{(i,j)}\}_{(i,j)\in\Phi_{m}}$.
If we perform IDFT for the subtraction 
$\{\tilde{I}_{(i,j)}-\breve{I}_{(i,j)}\}_{0\le i,j<q-1}$ (Fig. \ref{systematic}(j)), 
i.e., we compute
$$c_{(r,s)}:=\sum_{0\le i,j<q-1}
\left(\tilde{I}_{(i,j)}-\breve{I}_{(i,j)}\right)\alpha^{-ir-js},
$$
then $\{c_{(r,s)}\}$ is a code-word in $\mathcal{C}(m)$ since 
$c(\alpha^{i},\alpha^{j})=\tilde{I}_{(i,j)}-\breve{I}_{(i,j)}=0$ for 
$(i,j)\in\Phi_{m}$.
Moreover, it is systematic, as observed at Fig. \ref{systematic}(k),
and in fact we have 
$c_{(r,s)}=I_{(r,s)}$ for $(\alpha^{r},\alpha^{s})\in\wp'$ since the IDFT of 
$\breve{I}_{(i,j)}$ vanishes at $\wp'$ by Theorem \ref{lemma of Fourier}.

While the error-value estimation was performed by using the IDFT of (n) in 
\cite{Sakata-Jensen-Hoholdt}, it is efficiently incorporated into our 
procedure.
In the encoding of Section \ref{Encoding 2-D}, each procedure of encoding 
and decoding contains either the DFT or IDFT.
Although in this section, each step of encoding and decoding requires both the DFT and IDFT, we can use only one DFT calculator for all transforms in practical circuits for the encoder and decoder.

\begin{figure*}[!t]
\centering
  \resizebox{14.5cm}{!}{\includegraphics{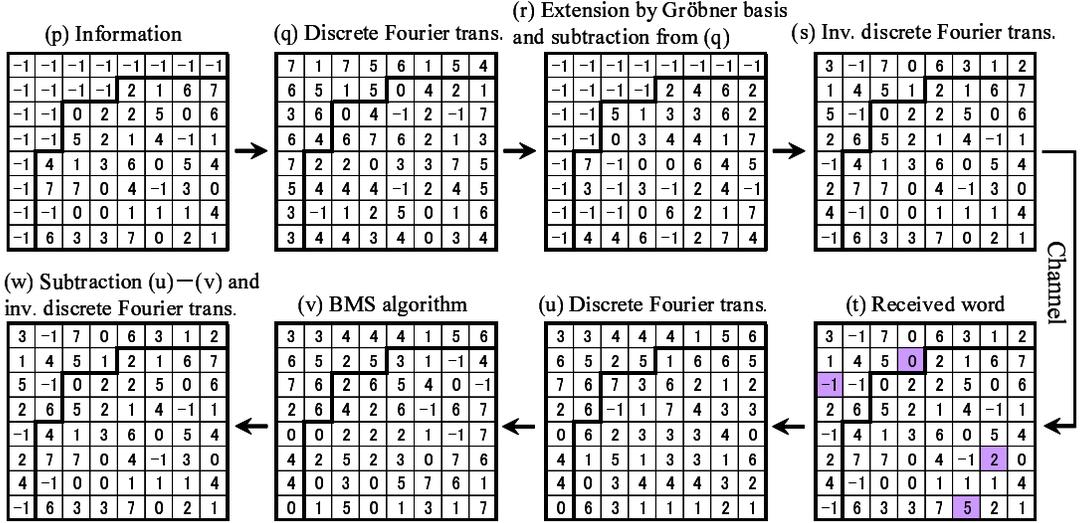}}
\vspace{-2mm}
\caption{Flow chart of systematic encoding by Gr\"obner basis, decoding by BMS algorithm with DFT for a hyperbolic cascaded RS code; array (s) represents a systematic code-word. The shaded values in (t) denote the values with errors added in the channel. Array (w) indicates that four errors have been corrected.\label{HCRS}}
\vspace{-3mm}
\end{figure*}

Recall that the systematic matrix-encoding described in Section \ref{Codes 
on algebraic curves} requires multiplications of the $k\times(n-k)$ matrix;
our method requires the calculators of the 2-D feedback shift-registers
and memory-elements for at most $a\times(n-k)$ coefficients,
which correspond to the $(n-k)$ coefficients of the generator polynomial for 
RS codes.

\section{Application to HCRS codes}

Our encoding and decoding scheme is widely applied to various algebraic 
codes such as 2-D cyclic codes and hyperbolic cascaded RS 
(HCRS) codes; for these codes,
the encodings in 
sections 3 and 4 are similarly performed
except for the total order in the BMS algorithm. 
Here, we deal with the systematic encoding of HCRS codes.

In this section let 
$\Phi_{m}:=\{(i,j)\in\mathbb{Z}_{0}^{2}\,|\,(i+1)(j+1)<m\}$.
A HCRS code 
\cite{Saints} over $K:=\mathbb{F}_{q}$ is defined as
$$
\mathcal{C}(m):=\left\{
(c_{r,s})_{0\le r,s<q-1}\left|\,
c(\alpha^{i},\alpha^{j})=0,\;(i,j)\in\Phi_{m}
\right.\right\},$$
where $c(x,y):=\sum_{0\le r,s<q-1}c_{r,s}x^{r}y^{s}$.
Then, the minimum distance $d$ of $\mathcal{C}(m)$ is bounded as $d\ge m$.
In Fig. \ref{HCRS}, $\mathcal{C}(9)$ over $\mathbb{F}_{9}$ is demonstrated 
for four-error correction.

The non-systematic encoding is equal to the IDFT of (p).
To encode systematically, we first compute a Gr\"obner basis of an ideal
$\{f\in R\,|\,f(\alpha^{i},\alpha^{j})=0,\;(i,j)\in\Phi_{m}\}$,
where $R:=K[x,y]/(x^{q-1}-1,y^{q-1}-1,)$,
with respect to a total order $(i,j)\prec(i',j')\Longleftrightarrow$
\begin{align*}
&(i+1)(j+1)<(i'+1)(j'+1),\quad\mathrm{or}\\
&(i+1)(j+1)=(i'+1)(j'+1)\;\wedge\;j<j'.
\end{align*}
The elements of the basis that is needed for the extension in the 
systematic encoding are shown below.
$$
\resizebox{6cm}{!}{\includegraphics{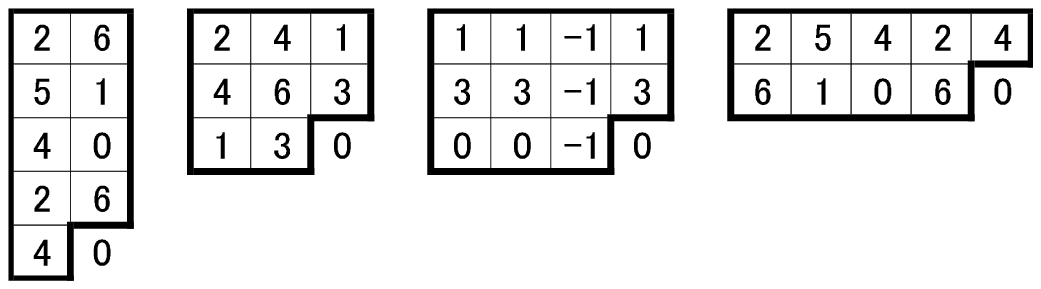}}
\vspace{-2mm}
$$
Then, the values after DFT on $\Phi_{m}$ in (q) are extended by the recurrence formula similar to \eqref{generation}.
Thus, the IDFT of (r), where (r) equals (q) minus the extended array, is a systematic code-word (s).

To decode a received word (t) from the channel, we perform, for syndrome 
values on $\Phi_{m}$ in (u), the BMS algorithm with respect to the total 
order ($\prec$) \cite{Saints}. In the case of our example, the error-locator 
polynomials are expressed as follows.
$$
\resizebox{4cm}{!}{\includegraphics{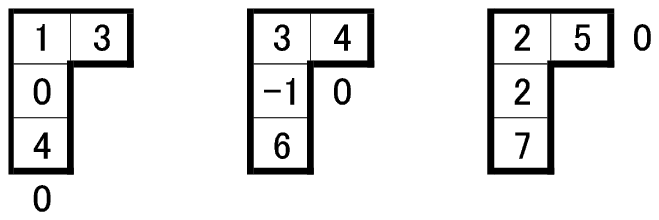}}
\vspace{-3mm}
$$
The recurrence similar to \eqref{generation} by the above basis
extends the syndrome values on $\Phi_{m}$ to the entire plane,
as Fig. \ref{HCRS}(v).
Finally, the IDFT of $(\mathrm{u}-\mathrm{v})$ in Fig. \ref{HCRS}
provides the correct transmitted word Fig. \ref{HCRS}(w).

\begin{figure*}[!t]
\centering
  \resizebox{15cm}{!}{\includegraphics{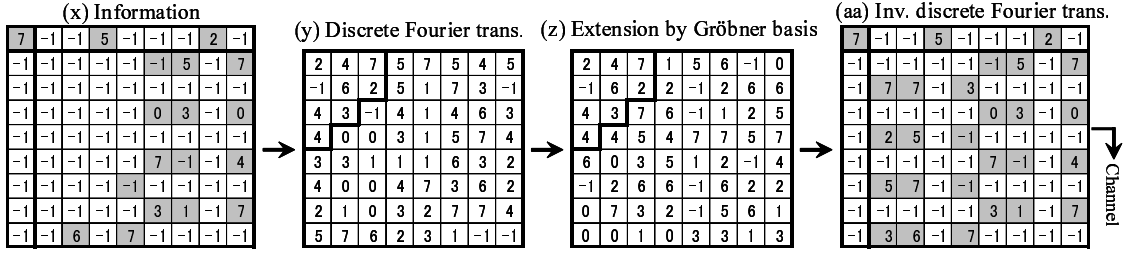}}
\vspace{-2mm}
\caption{Flow chart of systematic encoding by Gr\"obner basis and DFT for code $\mathcal{C}(11)$ on all finite GF(9)-rational points (including zero components) of the Hermitian curve $y^{3}+y=x^{4}$; array (aa) represents a systematic code-word in $\mathcal{C}(11)$.
\label{zero}}
\vspace{-4mm}
\end{figure*}

\section{The case of RS codes}\label{Motivation}
Recall the encoding for RS codes by Euclidean division:
\begin{equation}\label{Euclidean}
c(x):=I(x)-R(x)=Q(x)G(x),\:\deg(R)<n-k,
\end{equation}
where $I(x)=\sum_{0\le\kappa<n}I_{\kappa}x^{\kappa}$ is an information 
polynomial with $I_{\kappa}=0$ for $0\le\kappa<n-k$;
$G(x)=(x-1)\cdots(x-\alpha^{n-k-1})$, the generator polynomial;
$R(x)$, the remainder of the division with quotient $Q(x)$.
Then, it is apparent that $(c_{h})$ from $c(x)=\sum_{0\le h<n}c_{h}x^{h}$
is a code-word of the RS code
$$
\mathcal{C}(m):=\left\{
(c_{h})_{0\le h<n}\in \mathbb{F}_{q}^{n}\left|\,
c(\alpha^{i})=0,\,0\le i\le m
\right.\right\}
$$
with $n:=q-1$ and $m:=n-k-1$.
This method is {\it systematic}, i.e., $c_{h}=I_{h}$ for $n-k\le h<n$.

If we have received a polynomial $\overline{c}(x)=\sum_{0\le 
h<n}\overline{c}_{h}x^{h}=c(x)+e(x)$ containing an error polynomial
$e(x)$ in the channel,
the values of syndromes $\{e(\alpha^{\kappa})\}_{0\le\kappa<n-k}$ 
can be computed as $\{\overline{c}(\alpha^{\kappa})\}$ by substituting the 
roots of $G(x)$ into $\overline{c}(x)$. We notice that
$\overline{c}(\alpha^{\kappa})=
\sum_{0\le h<n}\overline{c}_{h}\alpha^{\kappa h}$
can be also considered to be the DFT of $\{\overline{c}_{h}\}$.
Thus, we obtain another encoding method ({\it non-systematic}) by the IDFT
$c_{h}:=\sum_{0\le i<n}I_{i}\alpha^{-ih}$.
Then, $(c_{h})_{0\le h<n}$ is another code-word of $\mathcal{C}(m)$ since
$$
c(\alpha^{i'})=
\sum_{0\le h<n}\sum_{0\le i<n}I_{i}\alpha^{-ih+i'h}=(q-1)I_{i'}.
$$

It is possible to systematically encode by using an alternative procedure.
From \eqref{Euclidean}, we obtain $I(\alpha^{\kappa})=R(\alpha^{\kappa})$ 
for $0\le\kappa<n-k$.
Moreover, we define array $(d_{h})_{0\le h<n}$ inductively by
$$
d_{h}:=\left\{\begin{array}{cl}
I(\alpha^{h})&0\le h<n-k,\\
-\sum_{i=0}^{n-k-1}G_{i}s_{i+h-(n-k)}&
n-k\le h<n,
        \end{array}\right.
$$
where $G(x)=\sum_{i=0}^{n-k-1}G_{i}x^{i}+x^{n-k}$.
Then, it follows that $d_{h}=\sum_{i=0}^{n-k-1}R_{i}\alpha^{ih}$ with
$R(x)=\sum_{i=0}^{n-k-1}R_{i}x^{i}$ not only for $0\le h<n-k$ but also for $n-k\le h<n$.
Thus, the IDFT $\left(-d(\alpha^{-i})\right)$ for $(d_{h})_{0\le h<n}$ is observed to agree with $(R_{i})$ of $R(x)$ by using Fourier inversion formula;
$c(x):=I(x)-R(x)$ again indicates the encoding,
and moreover we obtain two ways of calculating $R(x)$, i.e., a 
commutative diagram.
\vspace{-4mm}
\begin{figure}[!h]
\centering
  \resizebox{5cm}{!}{\includegraphics{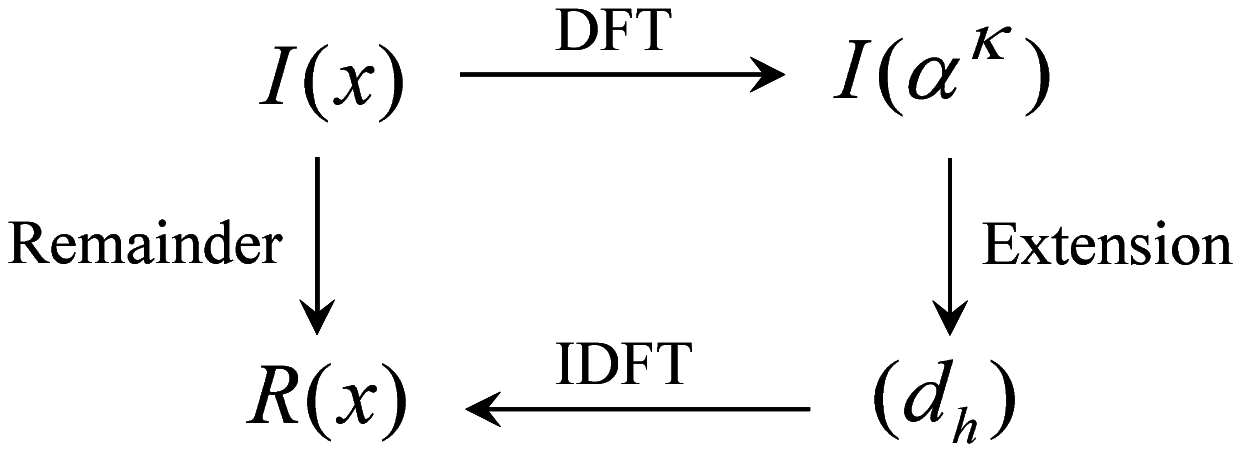}}
\vspace{-4mm}
\end{figure}

Thus, we obtain two encoding methods for RS codes, which we have generalized.

\section{Recursive formula from Gr\"obner basis}
\label{Proof of Theorem}

We generate $\{I_{(i,j)}\}$ for $0\le i,j<q-1$ recursively from 
$\{I_{(i,j)}\}_{(i,j)\in\Phi}$ as in the encoding at Section \ref{Encoding 
2-D} and \ref{Systematic encoding}, which is stated here more precisely. It 
may be assumed that we have the support $\Phi_{m}\subset\{(i,j)\,|\,0\le 
i<q-1,\,0\le j<a\}$ and that each Gr\"obner basis consists of $a+1$ 
elements $\{f^{(\iota)}\}_{0\le\iota<a}\cup\{g\}$, where 
$f^{(\iota)}=\sum_{(i,j)\in\Phi_{m}}f_{(i,j)}^{(\iota)}x^{i}y^{j}+x^{i_{\iota}}y^{\iota}$ 
with $i_{\iota}:=\max\{i+1\,|\,(i,\iota)\in\Phi_{m}\}$ and 
$g=\sum_{(i,j)\in\Phi}g_{(i,j)}x^{i}y^{j}+y^{a}$.
Then the recurrence for $(r,s)\not\in\Phi_{m}$
is defined as follows:
\begin{equation}\label{generation}
I_{(r,s)}\!:=\!\left\{\hspace{-1.5mm}
\begin{array}{cl}
-\sum_{(i,j)\in\Phi_{m}}
f_{(i,j)}^{(\iota)}I_{(i,j)+(r,s)-(i_{\iota},\iota)} &
s=\iota,\\
-\sum_{(i,j)\in\Phi_{m}}
g_{(i,j)}I_{(i,j)+(r,s)-(0,a)} &
s\ge a.
        \end{array}\right.
\end{equation}
The resulting values do not depend on the order of the generation because of 
the property of Gr\"obner bases.

\section{Treatment of locations including zero}\label{Treatment}
With regard to the systematic encoding in Section 4, we can treat locations 
including zero in a manner similar to the case of non-zero locations.
There are three 
$\mathbb{F}_{9}$-rational points $(-1,-1)$, $(-1,2)$, and $(-1,6)$ for our 
example of Hermitian codes, which are denoted by the shaded boxes in the top row of Fig. \ref{zero}(x).
Although we cannot compute the DFT for three information 
symbols at these locations, we note that if an error $1=\alpha^{0}$ has 
occurred on, e.g., $(-1,-1)$, then the syndrome values are all $-1$ except for $\alpha^{0}$ at $(0,0)\in\Phi$ since they are computed by the substitution 
of $(-1,-1)$ into $\{x^{i}y^{j}\}_{0\le i,j<q-1}$. We also note that the IDFT of $\{x^{i}y^{j}\}|_{(x,y)=(-1,-1)}$ equals an $8\times8$ all-$\alpha^{0}$ array. 
Thus the analogue of DFT is obtained for information 7 at $(-1,-1)$ as 
only 7 at $(0,0)\in\Phi$. Similarly, the analogue is obtained for
information 5 at $(-1,2)$ as only $[5,7,1,3,5,7,1,3]$ in the first row of 
$\Phi$, which also equals the one-dimensional (1-D) DFT for 
$[-1,-1,5,-1,\cdots,-1]$; for information 2 at $(-1,6)$,
the analogue is obtained as 
$[2,0,6,4,2,0,6,4]$ at the top, which is also the 1-D DFT for 
$[-1,\cdots,-1,2,-1]$.
Hence, we obtain the analogue of DFT for the array (x) 
as the array (y) by summing and obtain the redundant part of the 
code-word (aa) by the IDFT of the extended array (z).
Notice that the IDFT of $(\mathrm{y}-\mathrm{z})$ such as in Section 4 provides the sum of the parts of the code-word on $\Phi$ and the IDFT of the analogue arrays of DFT, i.e., all-$\alpha^{7}$ 
array, all-$(-1)$ array except $[1,\cdots,1]^{T}$ in the third column, which 
is the IDFT of $\{\alpha^{5}x^{i}y^{j}\}|_{(x,y)=(-1,2)}$, and all-$(-1)$ except $[6,\cdots,6]^{T}$ in the seventh column.

Thus, we have completed the treatment of locations including zero components.
\vspace{-1mm}

\section*{Acknowledgment}
\vspace{-1mm}
This work was partly supported
by the Academic Frontier Project
for Future Data Storage Materials Research
by the Japanese Ministry of Education, Culture, Sports,
Science and Technology (1999--2008),
Toyota Physical and Chemical Research Institute,
and Storage Research Consortium (SRC).

\end{document}